\documentclass[conference]{IEEEtran}
\IEEEoverridecommandlockouts
\usepackage{cite}
\usepackage{amsmath,amssymb,amsfonts}
\usepackage{algorithmic}
\usepackage{graphicx}
\usepackage{textcomp}
\usepackage{xcolor}
\def\BibTeX{{\rm B\kern-.05em{\sc i\kern-.025em b}\kern-.08em
    T\kern-.1667em\lower.7ex\hbox{E}\kern-.125emX}}

\usepackage{booktabs}

\usepackage[hidelinks]{hyperref}

\usepackage{listings}
\lstdefinelanguage{SQL}{
  morekeywords={SELECT,FROM,WHERE,INSERT,UPDATE,DELETE,CREATE,ALTER,DROP,
                TABLE,VIEW,INDEX,JOIN,ON,AS,AND,OR,NOT,IN,IS,NULL,TRUE,FALSE},
  sensitive=false,
  morecomment=[l]{--},
  morestring=[b]',
}
\begin{document}

\title{Towards Observation Lakehouses: Living, Interactive Archives of Software Behavior}

\author{\IEEEauthorblockN{Marcus Kessel}
\IEEEauthorblockA{\textit{University of Mannheim, Germany}\\
marcus.kessel@uni-mannheim.de, 0000-0003-3088-2166}
}

\maketitle

\begin{abstract}
Code-generating LLMs are trained largely on static artifacts (source, comments, specifications) and rarely on materializations of run‑time behavior. As a result, they readily internalize buggy or mislabeled code. Since non‑trivial semantic properties are undecidable in general, the only practical way to obtain ground-truth functionality is by dynamic observation of executions. In prior work, we addressed representation with Sequence Sheets, Stimulus-Response Matrices (SRMs), and Stimulus-Response Cubes (SRCs) to capture and compare behavior across tests, implementations, and contexts. These structures make observation data analyzable offline and reusable, but they do not by themselves provide persistence, evolution, or interactive analytics at scale. In this paper, therefore, we introduce observation lakehouses that operationalize continual SRCs: a tall, append‑only observations table storing every actuation (stimulus, response, context) and SQL queries that materialize SRC slices on demand. Built on Apache Parquet + Iceberg + DuckDB, the lakehouse ingests data from controlled pipelines (LASSO) and CI pipelines (e.g., unit test executions), enabling n‑version assessment, behavioral clustering, and consensus oracles without re‑execution. On a 509‑problem benchmark, we ingest $\approx$8.6M observation rows ($<$51MiB) and reconstruct SRM/SRC views and clusters in $<$100ms on a laptop, demonstrating that continual behavior mining is practical without a distributed cluster of machines. This makes behavioral ground truth first‑class alongside other run‑time data and provides an infrastructure path toward behavior‑aware evaluation and training.
The Observation Lakehouse, together with the accompanying dataset, is publicly available as an open‑source project on GitHub: \url{https://github.com/SoftwareObservatorium/observation-lakehouse}
\end{abstract}

\begin{IEEEkeywords}
testing, mining, oracle, evolution, behavior, analytics, repository, lakehouse, observations, scalability, database
\end{IEEEkeywords}

\section{Introduction}
\label{sec:introduction}

Most code-generating LLMs are trained on static artifacts -- source code, comments, and specifications -- rather than on materializations of run‑time behavior \cite{NEURIPS2024_6efcc7fd}. Such models inevitably internalize patterns from buggy or mislabeled code as readily as from correct implementations, because they are rarely, if ever, exposed to trustworthy evidence of functional run-time behavior (i.e., functionality) \cite{chen2021evaluatinglargelanguagemodels,10174227}. Improving reliability, therefore, requires exposing models to ground‑truth behavior captured under execution.

Rice's theorem rules out deciding non‑trivial semantic properties of programs in general \cite{riceTheorem}, so for real systems the only practical way of doing this is through dynamic observation -- that is, to execute software where it is expected to meet its advertised functionality and record what it actually does, either during controlled tests (i.e., software testing \cite{ammann2017introduction}) or during real usage (i.e., monitoring and tracing). Current technology makes this much harder than it should be. Unit‑testing frameworks typically discard most observations (observable outputs to calls), preserving only pass/fail outcomes, and burying stimuli in scaffolding code. Execution tracing captures the opposite extreme -- massive, low‑level call/state data -- from which functional responses must be mined laboriously. In short, testing captures too little, while (extensive) tracing captures too much, and both obscure the functionality that is important.

In earlier work, we addressed the representation gap by introducing three data structures tailored to the observation and recording of run‑time functional behavior \cite{kesselOS2023}: Sequence Sheets to encode the precise invocations and the responses they elicit, the Stimulus-Response Matrices (SRMs) to compare many implementations against many tests, and Stimulus-Response Cubes (SRCs) (formerly SRHs) to analyze ensembles of SRMs across additional dimensions such as repetitions, environments, and metrics. In an SRM, rows are tests and columns are implementations of the same functional abstraction (i.e., coding problem). Each cell stores the actuation (operation, inputs, outputs) for that test-implementation pair. SRCs bundle related SRMs into a multi‑dimensional structure for navigation and analysis. These structures make behavioral data analyzable offline through data-driven analytics, and reusable (replication, reproduction, repurposing) \cite{kesselOS2023}.

While the efficacy of these data structures has been demonstrated, also at large scales \cite{kesselOS2023,kesselNextGen2023}, their current implementation in the LASSO platform assumes semi-sanitized scenarios where the tests and implementations that populate SRCs are ``curated'' for specific test-driven software experiments, or for particular differential GAI applications \cite{kesselGAI2024}. At present, they are not supported by any infrastructure that can collect this information, at scale and at high speed, from mainstream development environments and execution platforms, and they were not developed to allow SRC data to live alongside all the other data that is generated during software execution. More specifically, SRCs currently lack an environment/infrastructure that i) keeps every actuation, not just pass/fail verdicts; (ii) evolves without friction as new tests, implementations, environments, measurements and context appear; and (iii) answers interactive queries that reconstruct SRC views and support higher-level analyses (e.g., behavioral clustering, n-version assessment) without a distributed cluster of machines, and (iv) allows SRC data to live alongside other forms of structured and unstructured data within a broader record of run-time software execution. In short, we need an infrastructure that realizes the vision of \textit{continual SRCs} -- motivated recently in our research agenda for behavior-aware LLMs \cite{kessel2025} -- that grow as code is tested and make it integrable in and accessible from the current software engineering data tool landscape.


To this end, we introduce the notion of an \textit{Observation Lakehouse} that operationalizes the continual SRCs concept. Realized as a Python application based on a modern data lakehouse architecture \cite{armbrust2021lakehouse}, it uses Apache Iceberg \cite{apache_iceberg} as a transactional table format over columnar Apache Parquet files \cite{apache_parquet_format}, with DuckDB for high-performance querying \cite{duckdb}. Our approach unifies behavioral observations from two complementary streams -- controlled software experiments and real-world CI pipelines -- into a single, append-only data model. This foundation makes large-scale differential analysis practical and efficient, allowing users to perform n-version assessment of code variants, use behavioral clustering to identify equivalence classes, and compute consensus oracles without re-executing historical runs. The feasibility of this approach is demonstrated on a benchmark of over 8 million observation rows, where complex analytical queries -- using SQL -- complete in under 100ms on a standard laptop with a storage footprint of less than 51MiB. This paper contributes a lakehouse infrastructure that provides a practical foundation for mining software behavior at scale, addressing a key obstacle to creating more behavior-aware GAI.

The remainder of the paper is organized as follows. Section~\ref{sec:architecture} details the lakehouse architecture and the data-modeling choices that enable a continual SRC. Section~\ref{sec:performance} reports ingestion throughput, query latency, and the cost of behavioral clustering that underpins consensus‑oracle construction. Section~\ref{sec:conclusion} discusses extensions (semantic search, MCP server, federation) and outlines current limitations, before concluding with a call to the research community to adopt and extend the observation lakehouse.

\section{Observation Lakehouse Architecture}
\label{sec:architecture}

In this section, we describe the design and implementation of the Observation Lakehouse, a platform that materializes the continual SRC \cite{kessel2025}. The architecture is designed to satisfy four core requirements: (1) support for concurrent, incremental ingestion of both experimental and practitioner data; (2) unrestricted schema evolution for dimensional extensibility; (3) efficient on-the-fly reconstruction of arbitrary SRMs; and (4) fast, expressive analytics over massive observation sets.


\subsection{Conceptual Model: A Continual and Unified SRC}

\begin{figure}
    \centering
    \includegraphics[scale=0.5, trim={0cm 0.5cm 12cm 0cm},clip]{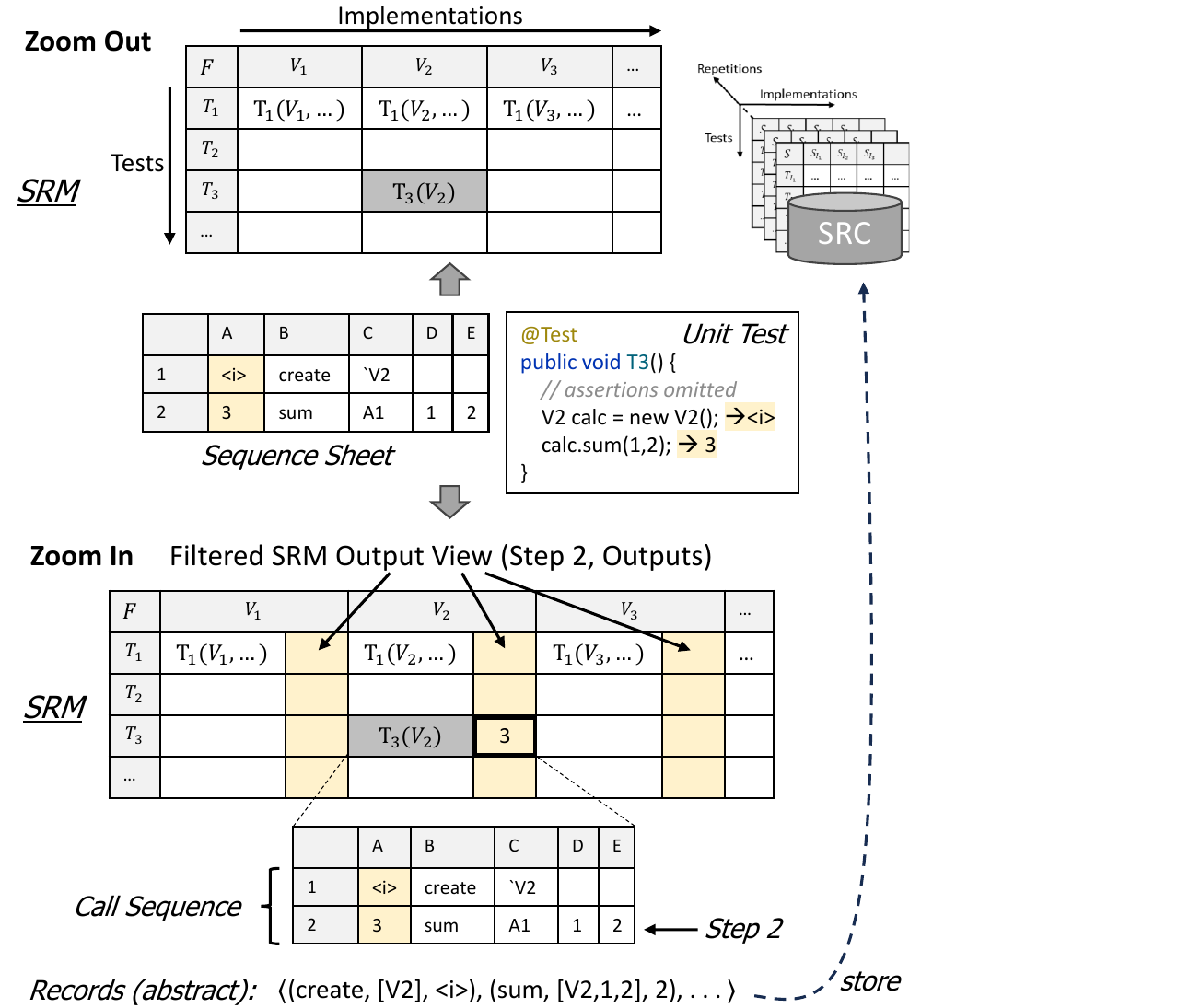}
    \caption{Conceptual Overview: SRC and SRMs - Zoom Out, In and Views}
    \vspace{-0.5cm}
    \label{fig:concepts}
\end{figure}

An SRC is a collection of SRMs (see Figure~\ref{fig:concepts}), one for each functional abstraction (either a coding problem described in some way or a concrete unit under test like a class). An SRM is a logical matrix where rows represent tests (stimuli) and columns represent code implementations. A crucial requirement is that this matrix must be continual: new tests (rows) and implementations (columns) can be added at any moment. Our platform realizes this through a flexible, \textit{append-only ingestion model}. New test executions are simply appended as new observation records, preserving the ``continual'' property without any costly data re-organization.

To support this, we define the atomic unit of observation as an invocation step record. This is a record capturing a single (operation, inputs, outputs) triple from an execution trace, along with its full context (e.g., implementation, test, and execution context like environment). A call-sequence is then an ordered list of these invocation step records:

  \[
    \langle (op_{1},\,in_{1},\,out_{1}),\;(op_{2},\,in_{2},\,out_{2}),\dots\rangle 
  \]

This abstraction is deliberately generic: an invocation step record can originate (1) from a sequence sheet used in test-driven experiments, or (2) from a mined unit-test method (e.g., from JUnit or Pytest), which we transform into this flat format.

While our lakehouse technically stores observations as a flat, granular stream of individual (un-nested) invocation step records, this serves as a base layer upon which various logical SRM views can be dynamically constructed for analysis. This crucial distinction provides significant analytical flexibility. Analysts can ``zoom in'' to a detailed, step-by-step call sequence for analyzing a single execution, effectively viewing a complete call sequence. They can create filtered views, such as an ``SRM output view'' that contains only the output values, which is ideal for n-version assessment and behavioral clustering.


\subsection{Architectural Foundation: A Data Lakehouse}

A conventional data lake (raw columnar files) lacks transactional guarantees, while a traditional RDBMS has a rigid schema and does not scale cost-effectively. To solve this, we employ a true data lakehouse architecture \cite{armbrust2021lakehouse} that combines the strengths of both. Our stack consists of ---

\begin{itemize}
    \item \textit{Columnar Storage (Apache Parquet \cite{apache_parquet_format})}: Provides efficient on-disk representation with high compression and predicate push-down (reduces how much data a query engine has to read from disk -- avoiding full scans).
    \item \textit{Transactional Table Format (Apache Iceberg \cite{apache_iceberg})}: Layered on top of Parquet, it provides ACID commits, snapshot isolation for time-travel, and, most importantly, schema evolution without rewriting existing files. This directly enables the SRC's ``dimensional extensibility'', allowing new columns to be added at any time.
    \item \textit{SQL Analytics Engine (DuckDB \cite{duckdb})}: Offers a familiar, high-performance query interface for interactive analysis. As an efficient, in-process, and vectorized execution engine, its tight integration with Apache Arrow \cite{apache_arrow} (an efficient in-Memory format) allows for zero-copy reads directly from Parquet files. This is ideal for providing near-instantaneous query feedback for analysts without the overhead of a distributed cluster.
\end{itemize}

Iceberg's atomic writes are critical, as they allow multiple ingestion pipelines to write concurrently to the lakehouse without race conditions.

\subsection{Data Model and Physical Design}

The logical SRC is materialized as three normalized Iceberg tables, mirroring a classic star-schema design known from data warehouses \cite{kimball2013data}. This three-table model, detailed in Table~\ref{tab:table-keys}, separates the core entities of a test execution -- the stimulus (i.e., test specification), the code unit under test, and the response. The primary advantage of this normalized design is massive de-duplication, as a single test or code implementation is stored only once.


\begin{table}[h]
\centering
\vspace{-0.1cm}
\caption{Tables and Key Columns}
\label{tab:table-keys}
\begin{tabular}{|p{2cm}|p{2cm}|p{3.4cm}|}
\hline
\textbf{Table} & \textbf{Contextual Keys} & \textbf{Core Columns \& Description} \\ \hline
code\_\newline implementations &
implementation\_id &
Stores source code and static metrics (e.g., LOC) for each implementation once. \\ \hline
tests &
test\_id &
Stores the stimulus definition (e.g., a sequence sheet or a mined unit‑test method) once. \\ \hline
observations &
impl\_id, test\_id, step\_id &
The ``fact'' table storing each invocation step of a call sequence: operation, inputs, outputs and context. \\ \hline
\end{tabular}
\end{table}

Two global identifiers, \texttt{data\_set\_id} and \texttt{problem\_id}, unify ingestion from heterogeneous sources. For experimental data, these map to a benchmark (e.g., ``HumanEval'') and a functional abstraction identifier (i.e., a coding problem). For practitioner data, they map to a repository (e.g., ``https://git/my-project/'') and a unit under test (e.g., fully-qualified name or path of a class).

A key data modeling choice is how to store additional dimensions (i.e., context), particularly dynamic measurements that apply to an entire test execution. Our normalized three-table model avoids duplicating large artifacts like source code and test definitions. However, for per-test-execution metrics like \texttt{branch\_coverage}, we employ a controlled denormalization strategy by repeating these values on every step-level record within a single test run. While this introduces data repetition, it is a deliberate design choice that is highly efficient in a columnar lakehouse. The underlying Parquet format uses compression techniques like run-length encoding (RLE) to store long runs of identical values with negligible storage overhead. This approach dramatically improves query performance by avoiding expensive JOINs to a separate ``test run metrics'' table, directly supporting the need for fast, interactive analysis.


The model is also designed to be programming language agnostic. The programming language of a unit under test or test specification is therefore modeled as a dimension, primarily realized as a column in the \texttt{code\_implementations} and tests tables. For analytical efficiency, this language attribute is also denormalized and stored with each step record, alongside other dimensions like the exact execution environment, following the same efficiency principles outlined above.

All three tables are physically partitioned on the two-level key (\texttt{data\_set\_id}, \texttt{problem\_id}). This is the most crucial design choice for performance. It ensures that all data related to a single logical SRM is co-located on disk \cite{duckdb_hive_partitioning}. When performing an analysis, the engine uses partition pruning to skip the vast majority of files and read only the relevant data, enabling efficient on-the-fly SRM reconstruction.


\vspace{-0.1cm}
\subsection{Illustrative Example: The Ingestion Flow}



To make our data model concrete, consider the end-to-end ingestion flow for a stateful Queue implementation, an example used to demonstrate the capture of operational semantics in the original SRC concept \cite{kessel2025}. First, a Java implementation of a queue is ingested into the \texttt{code\_implementations} table and assigned a unique implementation\_id (e.g., 'impl\_queue\_A'). Next, a test specification, with steps like create(), enqueue(1), and dequeue(), is ingested into the tests table and assigned a test\_id (e.g., 'test\_fifo\_1'). 

When the test is executed against the implementation, the core of our model comes into play. For each invocation step in the sequence, a single invocation step record is streamed into the observations table. For example, the execution of the enqueue(1) step would generate a record containing: contextual keys ('impl\_queue\_A', 'test\_fifo\_1'), call sequence information: step\_id: 2 (to preserve the execution order), and the stimulus-response triple ('enqueue', inputs: '{"value":1}', output: 'true'). This granular, linked model ensures that every observation is tied back to its originating syntactic code and stimulus, allowing our platform to reconstruct any logical SRM view on-the-fly for analysis.

\subsection{Implementation and Usage}

Our platform is realized in Python ($\ge$3.12) using \textsc{PyIceberg} and \textsc{PyArrow} for the ingestion and storage layers. The primary interface for data analysis is SQL, provided by DuckDB that has native support for Iceberg. This allows users to leverage a familiar, declarative language to perform rich behavioral analysis. Furthermore, the use of Arrow enables ``zero-copy'' data exchange with a large ecosystem of Python-based data science and machine learning libraries, facilitating seamless integration into existing workflows (e.g., analysts can use Jupyter notebooks together with DuckDB/Pandas/Polars).

\subsection{Analytical Capabilities Enabled by the Architecture}

The deliberate architectural choices of our lakehouse unlock a new class of behavioral analyses that are difficult or impossible with traditional software repositories. Since all historical observations are retained and efficiently queryable, we can perform rich analytics without needing to re-execute historical runs. 

\textit{N-Version Assessment and Differential GAI \cite{kesselGAI2024}}: Users can run n-version assessments of many alternative implementations, including LLM-generated variants, under identical tests. Our platform facilitates this by enabling behavioral clustering to identify equivalence classes and compute consensus oracles (majority behavior) for regression checks.

\textit{Behavioral Evolution Analysis}: By ingesting practitioner data with a git\_commit\_hash as a dimension, our lakehouse opens up new possibilities for behavioral evolution analyses -- by comparing code versions in SRM columns. Users can track the performance, correctness, and behavioral drift of specific code units over the course of a project's history, providing deeper insights into the semantic impact of code changes.

\vspace{-0.1cm}
\section{Performance}
\label{sec:performance}

This section presents a preliminary performance evaluation of the Observation Lakehouse, specifically assessing its ability to overcome the key scalability and feasibility obstacles that have hindered the creation of large-scale behavioral datasets. A key observation motivating this evaluation is that even a modest number of problems leads to a massive dataset; our workload, derived from just 509 coding problems, already results in $\approx$ 8.6M invocation step records (= rows in the observations table). This combinatorial explosion of behavioral data provides direct evidence for the need for platforms specifically designed to handle large data sets, justifying our choice of a data lakehouse architecture.

We evaluate the lakehouse along two axes: (1) the feasibility of ingesting this realistic, large-scale workload, and (2) the analytic efficiency of core queries required for SRM view reconstruction and behavioral analysis.

The workload is derived from 509 coding problems (taken from \cite{10103177}), a set of 13,384 LLM-generated implementations from four different LLMs, and a combination of 95,154 human-written and LLM-generated tests, yielding $\approx$8.6M step-level observations (see Table~\ref{tab:stats}). These observations were originally obtained using the LASSO platform \cite{lasso_arena}, a prototype system featuring an ``arena'' test driver for mass-execution. All measurements were performed on a laptop with an Intel Core i9, 64 GiB of RAM, and an NVMe SSD, running Ubuntu 24.04 and Python 3.12.



\begin{table}[ht]
\centering
\caption{Dataset and Table Characteristics}
\label{tab:stats}
\begin{tabular}{|p{2.5cm}|c|p{4cm}|}
\hline
\textbf{Metric}      & \textbf{Value}          & \textbf{Notes} \\
\hline
Observations & 8,556,455 & 95,154 call sequences (188.9 sequences per problem and 88.99 invocations per sequence) across 13,384 implementations (26.29 / problem) \\
\hline
observations size & 50.9MiB & On-disk size of the Parquet table \\
code\_implementations size & 9MiB & \\
test size & 9.5MiB &  \\
\hline
\textbf{Total Size}   & \textbf{69.4MiB}        & $\approx$ 8 MiB/1 million observations \\
\hline
\end{tabular}
\end{table}

The invocations in a call sequence include all invocations part of the entire test fixture (i.e., focal methods, test data initialization etc.). The high compression ratio -- 8.6M step records fitting into $\approx$51MiB (Table~\ref{tab:stats}) -- is a direct consequence of the columnar Parquet format and its efficient run-length encoding of repeated values, such as \texttt{implementation\_id}, within our normalized data model.

\subsection{Ingestion Performance}

Ingestion is performed in two phases: a one-time bulk import of all the code implementations and tests, which completes in $\approx$7 seconds, and the continuous ingestion of the observation records. We simulated the observation stream by writing the 8.6M records in batches using a Python worker with DuckDB and PyArrow. The total time to ingest all observations (including transformation costs) is $\approx$55s, achieving a sustained ingestion rate of $\approx$155,000 records per second, confirming the feasibility of populating the lakehouse from high-throughput sources like CI pipelines.

\subsection{Analytic Query Performance}

We evaluated three representative queries that map directly to core SRM workloads. Each query was run 10 times per problem on a cold cache, and we report the average per-problem latency in milliseconds.

\textbf{Q1: SRM Output View Reconstruction (Avg. Latency: $\approx$52ms)}: This query performs a partition-pruned scan of all steps for a specific problem (SELECT ... WHERE data\_set\_id=? AND problem\_id=?). Its sub-100ms latency is achieved because our (data\_set\_id, problem\_id) partitioning allows the DuckDB engine to read only the single, ``small'' partition on disk relevant to the SRM, ignoring all other data.

\textbf{Q2: Column-wise Behavioral Clustering (Avg. Latency: $\approx$29ms)}: This query identifies consensus behavior by grouping implementations based on their complete output trace. A behavioral fingerprint for each implementation is computed on-the-fly using a SQL CTE (WITH clause) to aggregate all outputs, and then grouped to find equivalence classes. The fact that this complex aggregation completes in under 30ms demonstrates the high performance of DuckDB's vectorized execution engine on in-memory Arrow data.

\textbf{Q3: Full Join for SRM Reconstruction (Avg. Latency: $\approx$90ms)}. This query performs a three-table join, combining observations with their corresponding tests and code\_implementations. The query remains highly performant because all tables are partitioned, limiting the join to a small subset of data -- validating the scalability of our normalized design.



The results confirm that our lakehouse design meets the key scalability requirements for a continual SRC. The high ingestion rate and efficient storage demonstrate its basic ability to handle large data sets. The sub-second latencies for on-the-fly SRM reconstruction and behavioral clustering encourage its suitability for interactive exploration and analysis of software run-time behavior.

\section{Future Work, Limitations, and Conclusion}
\label{sec:conclusion}

\paragraph{Future Work and Extensions}

Our tool provides a foundational platform for a continually evolving SRC, and we envision several key directions for future work. Our next major milestone is to ingest data from real-world projects via their CI pipelines. We have already prototyped lightweight driver extensions for JUnit5 and PyTest, demonstrating feasibility. We believe, this will enable large-scale behavioral evolution analyses, such as tracking semantic drift across commits. The tool's lightweight design also enables federation, where multiple lakehouses can be combined to form an ``organizational behavior memory'' or a global, community-driven SRC for large-scale experimentation. Finally, we are prototyping advanced analytical layers, including vector-based semantic search for similarity queries and a Model Context Protocol (MCP) server to enable ``augmented generation'' for LLMs (and agents).

The architecture is also inherently designed for evolution. The use of Iceberg not only enables the addition of new dimensions to the schema, such as code hashes for enhanced de-duplication, but also allows for the evolution of physical partitioning strategies to optimize for new query patterns as they emerge. This ensures that the platform can adapt to the future analytical needs of the community. 

\paragraph{Limitations}

Stimulus/response serialization is non-trivial for robust equivalence checking, especially for complex or non-JSON-friendly objects like binary streams. Our current JSON-based approach (cf. \cite{lasso_arena}) supports polyglot interoperability, but future work will need to evolve this protocol, likely by incorporating content-based hashing or specialized handlers to compare syntactically-different values (e.g., exceptions) as equivalent outputs. This necessitates the use of custom equivalence functions -- which could be realized as SQL User-Defined Functions.


\paragraph{Availability}

The implementation of our Observation Lakehouse is available as open source on GitHub. The repository includes the full source code, instructions and interactive notebooks to replicate the performance evaluation. The dataset used for our performance evaluation is available in \cite{kessel_2025_17791444}.

\paragraph{Conclusion}

This paper introduced the Observation Lakehouse, the first practical and scalable realization of the continual Stimulus-Response Cube (SRC) concept. By providing a true data lakehouse architecture, our tool enables the systematic mining of software execution behavior, unifying data from both controlled experiments and real-world CI pipelines. Our first performance evaluation demonstrates that the design is efficient and suitable for large-scale behavioral analysis. The Observation Lakehouse represents a critical step towards providing the community with a platform to curate, share, and analyze software behavior at large scale. We invite the research community to use and extend our tool in this open-scientific endeavor.

\bibliographystyle{IEEEtran}
\bibliography{IEEEabrv,literature}

\end{document}